\documentclass[letterpaper, 10 pt, conference]{ieeeconf}  
\IEEEoverridecommandlockouts           
\makeatletter
\let\NAT@parse\undefined
\makeatother

\usepackage[dvipsnames]{xcolor}

\newcommand*\linkcolours{ForestGreen}

\usepackage{subcaption}
\usepackage{times}
\usepackage{graphicx}
\usepackage{amssymb}
\usepackage{amsmath}
\usepackage{gensymb}
\usepackage{amsmath}
\usepackage{bbold}
\usepackage{breakurl}

\usepackage{url,hyperref}
\hypersetup{
colorlinks,
linkcolor=\linkcolours,
citecolor=\linkcolours,
filecolor=\linkcolours,
urlcolor=\linkcolours}

\usepackage{algorithm}
\usepackage{algorithmic}

\usepackage[labelfont={bf},font=small]{caption}
\usepackage[none]{hyphenat}

\usepackage{mathtools, cuted}

\usepackage[noadjust, nobreak]{cite}

\usepackage{tabularx}
\usepackage{amsmath}

\usepackage{float}

\usepackage{pifont}

\newcolumntype{Y}{>{\centering\arraybackslash}X}

\usepackage[]{placeins}

\usepackage{placeins}

\usepackage{tikz}

\usepackage[framemethod=tikz]{mdframed}

\usepackage{afterpage}

\usepackage{stfloats}

\usepackage{atbegshi}
\newcommand{\handlethispage}{}
\newcommand{\discardpagesfromhere}{\let\handlethispage\AtBeginShipoutDiscard}
\newcommand{\keeppagesfromhere}{\let\handlethispage\relax}
\AtBeginShipout{\handlethispage}

\usepackage{comment}

\title{\LARGE \bf
Quantum Phase Transition and Berry Phase in an Extended Dicke Model
}
\author{C. A. Estrada Guerra$^{1,2}$, J. Mahecha-G\'omez$^{1}$ and J. G. Hirsch$^{2}$
\thanks{$^{1}$ 
Universidad de Antioquia, Instituto de F\'\i sica, Facultad de Ciencias Exactas y Naturales, Calle 70 No. 52-21, Medell\'\i ­n, Colombia.
Email: alberto.estrada@udea.edu.co}%
\thanks{$^{2}$
Instituto de Ciencias Nucleares, Universidad Nacional Aut\'onoma de M\'exico, Apartado Postal 70-543, C.P. 04510 CDMX, Mexico.}%
}
\begin{document}

\maketitle
\thispagestyle{empty}
\pagestyle{empty}

\begin{abstract}
We investigate quantum phase transitions, quantum criticality, 
 and Berry phase for the ground state of an ensemble of  non-interacting two-level atoms embedded in a non-linear optical medium, coupled to a single-mode quantized electromagnetic field. The optical medium is pumped externally through a classical electric field, so that there is a degenerate parametric amplification effect, which strongly modifies the field dynamics without affecting the atomic sector.
Through a semiclassical description the different phases of this extended Dicke model are described.  
The quantum phase transition is characterized with
the expectation values of some observables of the system as well as the Berry phase and its first derivative, where such quantities serve as order parameters. 
It is remarkable that the model allows the control of the quantum criticality 
 through a suitable choice of the parameters of the non-linear optical medium, which could make possible the use of a low intensity laser to access the superradiant region experimentally.
\end{abstract}
\section{INTRODUCTION}

Thermal phase transitions occur in many physical systems. 
They are observed as changes in macroscopic properties, quite often 
discontinuous, when certain thermodynamic parameters of the system change.  Unlike thermal phase transitions, which happen at finite temperature, quantum phase transition (QPT) occurs at $T = 0$, where quantum fluctuations survive and are determined by the Heisenberg uncertainty principle. They are characterized by sudden changes in some order parameters (or their derivatives) when external parameters are varied. QPT is a topic of current interest in the areas of chaos, quantum optics, condensed matter, among others \cite{sachdev2007quantum,carr2010understanding,bastarrachea2015chaos, osterloh2002scaling}.

The Dicke model \cite{dicke1954coherence} describes the interaction between a single-mode electromagnetic field contained in an optical cavity and an ensemble of $N$  non-interacting identical atoms. The model is known in the literature of quantum optics and condensed matter due to superradiance; the emission process that interferes constructively and with an energy density proportional to $N^2$ \cite{gross1982superradiance}.  This system shows a second-order thermal phase transition, which occurs at finite temperature   \cite{hepp1973superradiant,wang1973phase}. Furthermore, for $T=0$, the ground state of the system becomes degenerate with the first excited state, when the atom-field interaction reaches
its critical value, leading the system from a normal to a superradiant phase. 
 In the thermodynamic limit, this level crossover becomes a real QPT, exhibiting a discontinuity in the derivatives of the ground state energy, the average number of photons and of excited atoms \cite{emary2003chaos}. QPT have been characterized in several extended Dicke models, adding 
 the interaction between the atoms \cite{li2006quantum,  chen2009interaction, li2013quantum, robles2015ground,  rodriguez2018critical}, nonlinear light interactions \cite{rodriguez2010quantum, guo2011quantum}, atom-optomechanical systems \cite{zhu2019entanglement}, and others. These studies generated interesting theoretical advances, but the experimental observation of the superradiant phase in a cavity-atom system remains challenging.
 It is due to the strong coupling that it is required between the atoms and photons, which must be of the order of the atomic and field frequencies \cite{kirton2019introduction}. Although the no-go theorem \cite{rzazewski1975phase,rzazewski1976thermodynamics,bialynicki1979no,knight1978super} prohibits the transition to the superradiant phase, several experimental results have shown that it is possible to reach that phase \cite{dimer2007proposed,nagy2010dicke,baumann2010dicke}.

From a different perspective, Berry showed that for  Hamiltonians which depend on a set of parameters that vary cyclically and adiabatically over time, the associated wave function acquires a phase factor of geometric nature, in addition to the dynamical phase due to temporal evolution \cite{berry1984quantal}. Therefore, since the ground state of many-body systems shows crossover or avoided crossover between the ground state and the first excited state, due to the variation of an external parameter of the system, the geometric phase and its derivative with respect to the external parameter can detect these  irregularities \cite{hamma2006berry}. The behavior of the geometric phase in the thermodynamic limit has been studied in the $XY$ spin model \cite{carollo2005geometric}, the Lipkin-Meshkov-Glick model \cite{cui2006geometric}, the Dicke model \cite{plastina2006scaling, chen2006critical,li2013quantum}, and experimentally in the Heisenberg $XY$ model \cite{peng2010observation} and the Zak phase in topological Bloch bands \cite{atala2013direct}.

In this work, we consider an extended model based on an ensemble of non-interacting two-level atoms, which are embedded in a nonlinear optical material pumped by a classical field.  Using semi-classical analysis, exact in the thermodynamic limit, it is shown that the critical value of the atom-field interaction 
can be noticeable reduced by the presence of the the non-linear terms, in comparison with the one needed  
in the standard Dicke model to reach the ultra-strong coupling regime.  We also show that
the expectation values of the field operators change, while the expectation values of atomic operators are not affected by non-linear terms. Finally, we derive the geometric phase of the ground state induced by the change in the quantized field, showing its usefulness in detecting the QPT, as well as the scaling behavior in the vicinity of the separatrix between phases.

The structure of the article is as follows. In section 2, we give an introduction to the Dicke model, showing their main characteristics. Afterwards, we present the extended model. The semiclassical description is given in section 3, where  we obtain the classical Hamiltonian, the classification of the fixed points, and the expectation values of both the atomic and photonic operators. In section 4 we deduce the expression of the Berry phase for the ground state, its derivative
with respect to the atom-field coupling parameter and its scaling behavior. Finally, the conclusions of our results are given in section 5.
\section{Model Hamiltonian}
The Dicke model describes an ensemble of $N$ non-interacting identical two-level atoms with atomic transition frequency $\omega_0$, interacting 
with a single-mode radiation field with frequency $\omega_f$ inside of a high finesse optical cavity. To obtain the Dicke model Hamiltonian, the following assumptions are made: $(i)$ the dipole approximation (long-wavelength limit, where the field wavelength is much greater than the size of region in which the atoms are confined). $(ii)$ Only two atomic levels that interact with the electromagnetic field are considered. $(iii)$ The quantum state of lowest energy is metastable so that we can neglect decays towards other atomic states. In this way, the Dicke model Hamiltonian is given by (taking $\hbar=1$ henceforth) \cite{emary2003chaos}
\begin{equation}
    \hat{H}_{D}=\omega_f \hat{a}^{\dagger}\hat{a}+\omega_0\hat{J}_z+\frac{\gamma}{\sqrt{N}}\left(\hat{a}^{\dagger}+\hat{a}\right)\left(\hat{J}_++\hat{J}_-\right),
\end{equation}
where $\gamma$ is the coupling constant for the atom-field interaction, $\hat{a}^{\dagger}$ and $\hat{a}$ are the creation and annihilation operators for the single-mode of the cavity, respectively, and satisfy the commutation rule $\left[\hat{a},\hat{a}^{\dagger}\right]=1$. The atomic ensemble is described through the pseudo-spin collective operators 
$\hat{J}=\sum_{k=1}^N \hat{j}_k$, with $\hat{j}_k = \hat\sigma_k/2$ is the kth component of the pseudo-spin operator for the atom $k$, which 
satisfy the SU(2) commutation relations $[\hat{J}_z,\hat{J}_{\pm}]=\pm \hat{J}_{\pm}$ and $[\hat{J}_+,\hat{J}_-]=2\hat{J}_z$, where $\hat{J}_z$, $\hat{J}_{\pm}$  are the atomic relative population and the atomic transition operators, respectively. The total spin quantum number is 
selected to be $j=N/2$, corresponding to the subspace which includes the ground state of the system and is completely symmetric.

In the Dicke model there are two types of phase transitions: $(i)$ a thermal second-order phase transition found by Hepp and Lieb \cite{hepp1973superradiant}, and mathematically described by Wang and Hioe \cite{wang1973phase}. When $\gamma>\sqrt{\omega_f \omega_0}$, there is a thermal
phase transition for a temperature $T_c$. For $T> T_c$, the system is in a normal phase where there are no atomic or photonic excitations, while for $T <T_c$ the system reaches the superradiant phase. $(ii)$ At $T=0$, there is a second-order QPT which occurs at the critical point $ \gamma_c^D = \sqrt{\omega_f \omega_0}/2 $. The system is in the normal phase, where the ground state is non-degenerate, and there are no atomic or photonic excitations for $ \gamma \leq \gamma_c^D $. For $ \gamma >\gamma_c^D $, the system is in the superradiant phase, where the symmetry is broken, causing degeneration in the ground state, and the photons and the atomic ensemble have macroscopic occupations \cite{emary2003chaos}.

In this work, we study an extended Dicke model (EDM), where a non-linear optical material is within a high-finesse optical cavity. It contains a quantized field mode and it is pumped with an electromagnetic field with frequency $2\omega_f$, described in the parametric approximation \cite{article}. This system is modeled through the inclusion of two terms containing non-linear operators, which correspond to the real and imaginary parts of the square of the
field amplitude, introduced by Hillery \cite{hillery1987squeezing}. In this way, a nonlinear effect known as degenerate parametric amplification (DPA) is produced within the cavity.  The Hamiltonian for EDM is
\begin{align}
    \hat{H}_{ED}= &\, \omega_f \hat{a}^{\dagger}\hat{a}+\omega_0 \hat{J}_z+\frac{\gamma}{\sqrt{N}}\left(\hat{a}^{\dagger}+\hat{a}\right)\left(\hat{J}_++\hat{J}_-\right) \nonumber \\
    & + \frac{K_1}{2}\left(\hat{a}^{\dagger \ 2}+\hat{a}^2\right)+i \frac{K_2}{2}\left(\hat{a}^{\dagger \ 2}-\hat{a}^2\right),
\label{EDM_Ham}
\end{align}
with $K_1$ and $K_2$ are the coupling of real and imaginary parts of the squared amplitude, respectively. Furthermore, the EDM Hamiltonian retains the same symmetry properties as Dicke model. When $K_1 = K_2 = 0$, we recover the Dicke Hamiltonian. Some properties for the case of $K_2 = 0$ have been studied in \cite{rodriguez2010quantum}.
When there is no interaction between the atoms and the field $(\gamma = 0)$, the Hamiltonian of the system is equivalent to a degenerate parametric amplifier with an extra term given by the energy of the atoms. 

Parity symmetry is characterized by
the unitary transformation $\hat{U}(\Phi)=e^{i\Phi \hat{\Lambda}}$, with $\hat{\Lambda}=\hat{a}^{\dagger}\hat{a}+\hat{J}_z+\sqrt{\hat{J}^2+1/4 \ \mathbb{1} }-1/2\ \mathbb{1}$ being the operator representing the total number of excitations, and its respective eigenvalues given by $\Lambda=n+m+j$.  The number of photons is represented by $n$,  and $m+j$ is the number of atomic excitations. This transformation acts on the atomic and photonic operators in the form $\hat{U}\hat{J}_+\hat{U}^{\dagger}=e^{-i\Phi}\hat{J}_+$ and $\hat{U}\hat{a}\hat{U}^{\dagger}=e^{i\Phi}\hat{a}$. Thus, the transformed Dicke Hamiltonian  is
\begin{align}
    \hat{U}\hat{H}_{ED}\hat{U}^{\dagger} & =\omega_f \hat{a}^{\dagger}\hat{a}+\omega_0\hat{J}_z+\frac{\gamma}{\sqrt{N}}\left(\hat{a}^{\dagger}\hat{J}_-+\hat{a}\hat{J}_+\right)\nonumber\\
    & +\frac{\gamma}{\sqrt{N}}\left(e^{-2i\Phi}\hat{a}^{\dagger}\hat{J}_++e^{2i\Phi}\hat{a}\hat{J}_-\right) \nonumber \\
    & + \frac{K_1}{2}\left(e^{-2i \Phi}\hat{a}^{\dagger \ 2}+e^{2i \Phi}\hat{a}\right)\nonumber \\ 
    & +i\frac{K_2}{2}\left(e^{-2i \Phi}\hat{a}^{\dagger \ 2}-e^{2i \Phi}\hat{a}\right).
\end{align}
The  Hamiltonian is invariant under the action of $\hat{U}(\Phi)$
for $ \Phi = 0, \pi $, and the invariant group is given by $ \mathcal{C}_2 = \{ \mathbb{1},e^{i\pi \hat{\Lambda}}\} $. Two projection operators emerge,  $\hat{P}_{\pm}=(\mathbb{1}\pm e^{i\pi \hat{\Lambda}})$, that classify the eigenvalues of $ \hat{\Lambda} $ into even $(+)$ and odd $(-)$. 
It implies that parity is a conserved quantity of the EDM,
 $\left[\hat{H}_D,\hat{U}(\pi)\right]=0$. 
 States belonging to each parity subspaces are not mixed by the Hamiltonian with states in the other subspace \cite{Castanos2011}.  

\section{Semiclassical Analysis}

The semiclassical version of the EDM Hamiltonian is obtained employing coherent states \cite{zhang1990coherent, nahmad2013mathematical}, both for the atomic sector (Bloch coherent states) and for the photonic sector (Glauber states), given, respectively, by
\begin{align}
    |z\rangle & = \frac{1}{(1+|z|^2)^j}e^{z\hat{J}_+}|j,-j\rangle, \label{Bloch_CS} \\
    |\alpha\rangle & = e^{-|\alpha|^2/2} e^{\alpha \hat{a}^{\dagger}}|0\rangle,
\label{Glauber_CS}
\end{align}
where $z$ and $\alpha$ are complex parameters. $\alpha$ depends on the  canonical variables for the electromagnetic field, $q$ and $p$, defined as $\alpha= (q+ip)/\sqrt{2}$. And  $z=\tan(\theta/2)e^{i\phi}$, where the angle $\theta$ 
measures the zenith angle with respect to the $-\hat{z}$ axis such that $j_z=-j \cos \theta$, and $\phi$ is the azimuthal angle. The canonical variables $(q, p)$ represent the classical analog of the electromagnetic field quadratures. In this form, the semiclassical Hamiltonian is 
\begin{align}
\mathcal{E} & = \mathcal{H}(q,p,\phi,\theta)  = \langle z| \otimes \langle \alpha| \hat{H}_{ED}|\alpha \rangle \otimes |z\rangle \nonumber \\
 & = \frac{\omega_f}{2}(q^2+p^2)-\omega_0 \, j\cos \theta + 2\sqrt{j}\, \gamma \, q \sin \theta \cos \phi \nonumber \\
 & + \frac{K_1}{2}(q^2-p^2)+K_2 \, q \, p.
\end{align}
The Hamilton's equations are given by
\begin{align}
    & \dot{q}=\frac{\partial \mathcal{H}}{\partial p}  = \omega_f p -K_1 p+K_2 q, \\
    & \dot{p}= - \frac{\partial \mathcal{H}}{\partial q}  = -\omega_f q-2\sqrt{j}\gamma \sin \theta \cos \phi-K_1 q-K_2 p,\\
    & \dot{\phi}=\frac{\partial \mathcal{H}}{\partial \theta}=\omega_0 j \sin \theta +2\sqrt{j}\gamma q \cos \theta \cos \phi, \\
    & \dot{\theta}=-\frac{\partial \mathcal{H}}{\partial \phi}=2 \sqrt{j}\gamma q \sin \theta \sin \phi.
\end{align}
In order to find the equilibrium points, we calculate $\nabla \mathcal{H}(q_c,p_c,\phi_c,\theta_c)=0$. The resulting critical points are $(q_c, p_c,\phi_c, \theta_c)=(0,0,\phi,0)$ and $(q_c, p_c,\phi_c, \theta_c)=(0,0,\phi,\pi)$ for any value of the coupling constant $\gamma$.  
The value $\theta_c=0$ corresponds to the South Pole of the Bloch sphere. It is stable for $\gamma \leq \gamma_c$  and unstable for $\gamma > \gamma_c$. The point $\theta_c=\pi$ represents the North Pole, which is unstable for any value of $\gamma$. In both of them the value of $\phi_c$ is not well defined. 
Two other critical points appear for $\phi_c = 0$ with $(q_c, \ p_c, \ 0,\ 
\arccos[(\gamma_c/\gamma)^2])$
and $\phi_c = \pi$ with $(-q_c,\ -p_c,\ \pi,\ 
\arccos[(\gamma_c/\gamma)^2])$.
 They are stable equilibrium points, 
 which can only exist for $\gamma \geq \gamma_c$. 
The quantities $q_c, \ p_c$ and $\gamma_c$ are given by
\begin{align}
    & q_c  =-\frac{\omega_0 \sqrt{j}\Gamma}{2\gamma_c}\sqrt{1-\Gamma^{-4}}, \label{qc}\\
    & p_c=-\frac{K_2}{\omega_f-K_1}q_c, \label{pc}\\
    & \gamma_c = \frac{1}{2}\sqrt{\frac{\omega_0(\omega_f^2-K_1^2-K_2^2)}{\omega_f-K_1}}, \label{gc}
\end{align}
with $\Gamma=\gamma/\gamma_c$ and $\gamma_c$ is the critical value of the atom-field coupling.
Enforcing $ \gamma_c $ to be real restricts the values of the $ (K_1, K_2) $ to be in the region $K_1<\omega_f$ and  $K_2<\sqrt{\omega_f^2-K_1^2}$. 
 As one the goals of this work is to find situations in which  $\gamma_c < \gamma_c^D$, the case $K_1>\omega_f$ will no be considered. It is worth to mention that, when $\gamma=0$, there is a rich semi-classical dynamics of the Hamiltonian \ref{EDM_Ham} which can be unveiled using SU(1,1) coherent states \cite{jarrett2007optical}.
 
When $K_1<\omega_f$,  it is clear from Eq. \ref{gc} that along the circular line $K_1^2+K_2^2=\omega_f^2$ the critical value is null: $\gamma_c=0$. 
The ground state undergoes a second-order phase transition and it becomes degenerate with the first excited state. When $K_1^2+K_2^2 \geq \omega_f^2$ the system is a the normal phase, characterized by the absence (in average) of excited atoms and photons within the cavity. When $K_1^2+K_2^2 < \omega_f^2$  the system is in the superradiant phase, with a macroscopic population of photons and excited atoms  inside the cavity \cite{gerry1988ground,gerry1990classical}.
It is interesting to notice that having non-linear materials inside the cavity which can fulfill this condition, the system will always be in the superradiant phase, for any positive value of the coupling parameter $\gamma$, however small it could be.  This represents an alternative way to achieve the strong coupling limit. 

\begin{figure}[H]
\centering
\includegraphics[width=0.7\columnwidth]{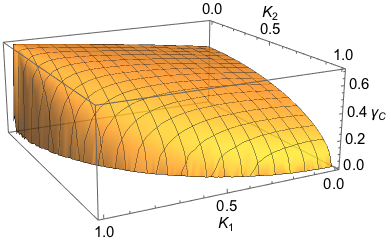}
\caption{Atom-field coupling constant $ \gamma_c $ as a function of $ K_1 $ and $ K_2 $, with resonance condition $ \omega_f=\omega_0=1$.}
\label{plot:critical_gamma}
\end{figure}

Figure \ref{plot:critical_gamma}  shows $ \gamma_c $ as function of the non-linear parameters, at
 the resonance condition, $ \omega_f=\omega_0=1 $,  in the region $0 \leq  K_1, K_2 \leq 1$.
  
The ground state scaled energy for each phases is 
\begin{equation}
    \epsilon_0=\frac{\mathcal{E}}{\omega_0 j}=\left\{
    \begin{array}{ll}
        -1, & \mbox{for }  \gamma \leq \gamma_c,\\
        -\frac{1}{2}\left(\Gamma^2+1/\Gamma^2\right), &   \mbox{for} \ \gamma > \gamma_c.
    \end{array}
    \right.
\end{equation}  

The functional dependence of the ground state energy on the coupling parameter $\gamma$ is the same as in the standard Dicke model \cite{Castanos2011b}. The influence of the non-linear terms $K_1, K_2$ is hidden inside the critical parameter  $\gamma_c$, given in Eq. \ref{gc}.

\begin{figure}[H]
\centering
\includegraphics[width=1\columnwidth]{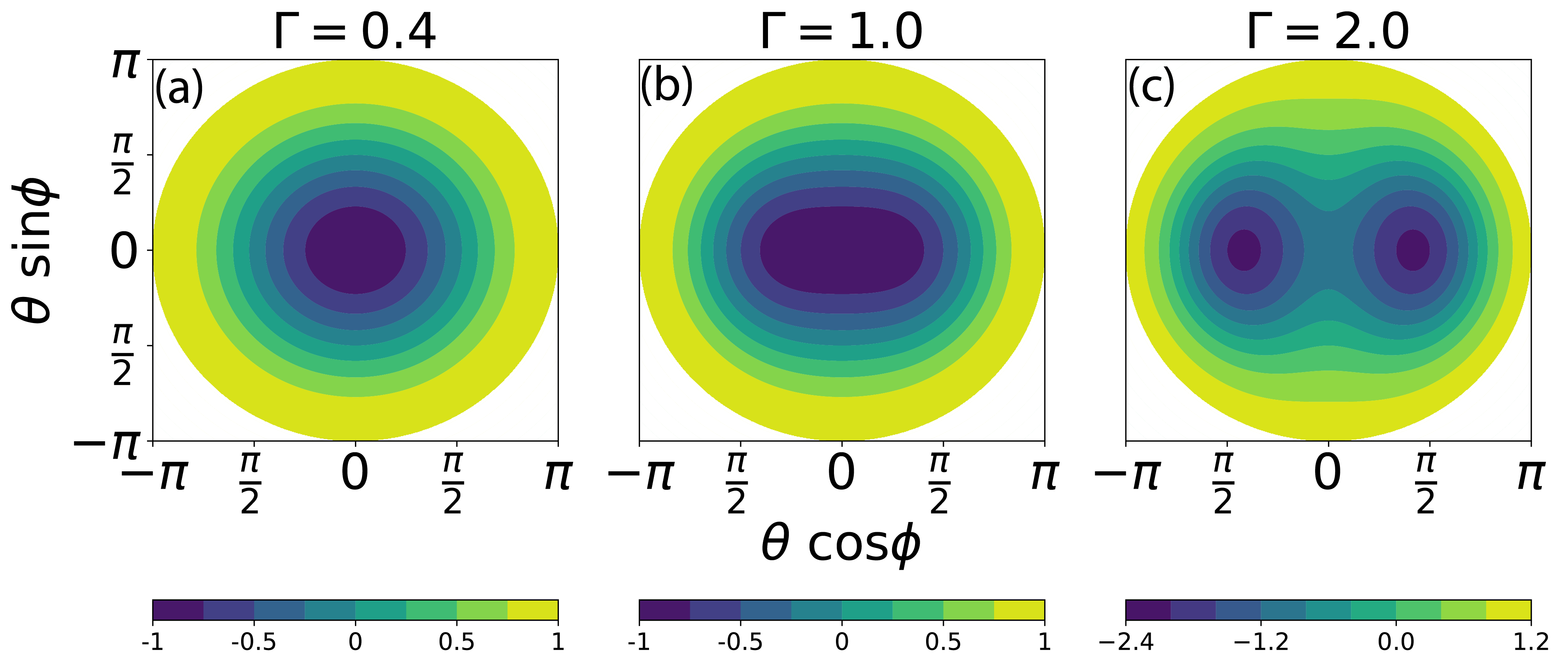}
\caption{Contours of the energy surfaces for Eq.  (\ref{energy_surface_theta_phi}) with $\Gamma = 0.4$, $\Gamma = 1.0$ and $\Gamma = 2.0$. In (a) and (b), the system is in the normal phase, and there is one stable minimum energy. In (c), a degeneration associated with symmetry breaking appears for $\theta=0,\pi$, and the system is in the superradiant phase.}
\label{plot:energy_surface}
\end{figure}

Another way to visualize the QPT is by expressing  the Hamiltonian in terms of the variables $\theta_c$ and $\phi_c$, that is, 
\begin{align}
\epsilon_0 & = -\cos \theta_c-\frac{\Gamma^2}{2} \sin^2 \theta_c \cos^2 \phi_c.  
\label{energy_surface_theta_phi}
 \end{align} 
In Figure \ref{plot:energy_surface}, we plot the contour plots of the ground state scaled energy showing the changes in the surface for three different values of $\Gamma$. In (a), we have  $ \Gamma = 0.4$, for which it is observed that $\theta = 0$ is the minimum stable of the energy $\epsilon_0$, and it belongs to the normal phase. In (b),  $ \Gamma = 1.0 $, and the equilibrium point is still stable, but with a stable region greater than the previous case. In (c), for $ \Gamma = 2.0 $, the energy of the system is doubly degenerated, for $\phi=0$ and $\phi=\pi$, and a saddle point appears. The system is in the superradiant phase.
The expectation values of the operators  $ \hat{q}, \ \hat{p}, \ \hat{a}^{\dagger} \hat{a} $ and $ \hat{J}_z $ serve as order parameters and characterize the quantum phases of the system. 
These quantities are calculated in the same way as the energy surface, using the coherent states. Table \ref{mean_values_and_fluctuations} shows these expressions and their respective fluctuations. Values in the normal phase can be obtained taking $ \Gamma \rightarrow 1 $. %
\begin{table}[H]
\centering
\resizebox{8.6cm}{!} {
\begin{tabular}{c c c}
\hline

\hline
Operator & Mean Value & Fluctuation \\
\hline
$\langle \hat{q}\rangle$& $-\frac{\omega_0\sqrt{j}\Gamma}{2\gamma_c}\sqrt{1-\Gamma^{-4}}$ & $\frac{1}{2}$ \\ 
$\langle \hat{p}\rangle$ & $-\frac{K_2}{\omega_f-K_1} \langle \hat{q}\rangle$ & $\frac{1}{2}$\\
$\langle \hat{a}^{\dagger}\hat{a}\rangle$ & $\frac{(\omega_f-K_1)^2+K_2^2}{2(\omega_f-K_1)^2}\langle \hat{q}\rangle ^2$ &$\frac{(\omega_f-K_1)^2+K_2^2}{2(\omega_f-K_1)^2}\langle \hat{q}\rangle ^2$ \\
$\langle \hat{J}_x \rangle$  & $j \sqrt{1-\Gamma^{-4}}$ & $\frac{j}{2}\Gamma^{-4}$ \\
$\langle \hat{J}_y \rangle$  & $0$ & $\frac{j}{2}$ \\
$\langle \hat{J}_z \rangle$  & $-j \Gamma^{-2}$ & $\frac{j}{2}(1-\Gamma^{-4})$\\
\hline
\end{tabular}
}
\caption{Expectation values of photonic and atomic operators in the superradiant phase with their respective fluctuations.}
\label{mean_values_and_fluctuations}
\end{table} 

\begin{figure}[h]
    \centering
     \includegraphics[width=1\columnwidth]{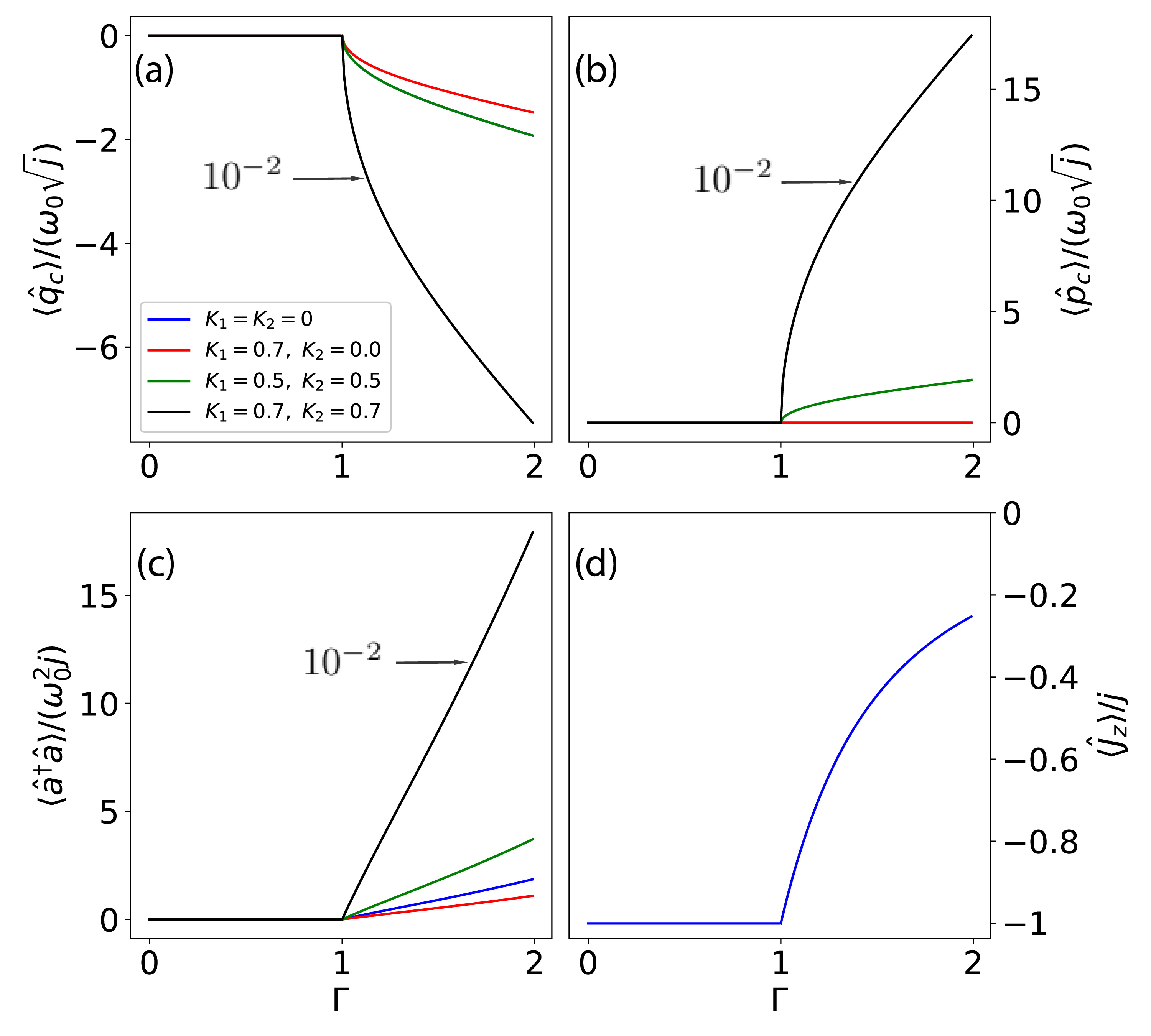}
    \caption{ $\langle \hat{q} \rangle/(\omega_0 \sqrt{j})$ and $\langle\hat{p} \rangle/(\omega_0 \sqrt{j})$,  $\langle\hat{a}^{\dagger}\hat{a}\rangle/(\omega_0^2 j)$ and  $\langle\hat{J}_z\rangle/j$ as a function of $\Gamma$ in the normal and superradiant phases for the ground state with different values of $K_1$ and $K_2$. The plots for the values $K_1=K_2=0.7$  are multiplied for $10^{-2}$. We take $\omega_f=1$.}
    \label{fig:mean_values_op}
\end{figure}

Their dependence on the nonlinear parameters $K_1$ and $K_2$ is shown in Figure \ref{fig:mean_values_op}, for different values of $(K_1, K_2)$. 
Notice that, when $K_1^2 + K_2^2$ approach $\omega_f^2$, the value of $\gamma_c$ goes to zero, $\Gamma=\gamma/\gamma_c$ grows without limits and tends to diverge for any finite value of $\gamma$. In this case the expectation of $q_c$, Eq. \ref{qc}, becomes approximately proportional to $\Gamma$, and also $p_c$, Eq. \ref{pc}, and the average number of photons in the cavity. This is the most significative effect of the presence of the non-linear material inside the cavity.  To account for this effect and allow a comparison, the values for $ K_1 = K_2 = 0.7 $ (black line) have been multiplied by $ 10^{-2} $. In the normal phase the operators $\langle \hat{q}\rangle, \langle \hat{p}\rangle$ and $\langle\hat{a}^{\dagger}\hat{a}\rangle$ have null expectation values.

In Figure \ref{fig:mean_values_op}(a), we plot $\langle \hat{q} \rangle/(\omega_0 \sqrt{j})$ for different values of $ K_1 $ and $ K_2 $. In the superradiant phase, the plots for $K_1=K_2=0$ (Dicke model - blue line) and $(K_1,K_2)=(0.5,0.5)$ (green line) are equals since they have the same value of $\gamma_c=0.5$. For $(K_1,K_2)=(0.7,0)$ (red line), with $\gamma_c=0.65$, the (absolute) values are larger than those of Dicke model. For $(K_1,K_2)=(0.7,0.7)$ (black line),  with $\gamma_c=0.13$, the expectation values are increased by two order of magnitudes due to the presence of the non-linear terms.
In Figure \ref{fig:mean_values_op}(b), we plot $\langle \hat{p} \rangle/(\omega_0 \sqrt{j})$. In the superradiant phase, for $K_1=K_2=0$ and $(K_1, K_2) = (0.7,0)$, the value of the quadrature $p$ is zero, since it is proportional to $K_2$, like in Dicke model \cite{bastarrachea2014comparative}. For  $(K_1,K_2)=(0.5,0.5)$, the value does not vanish, since $K_2$ is different from zero, even having the same value $ \gamma_c $ as Dicke model. When $(K_1,K_2)=(0.7,0.7)$, we find a much higher values than in the previous case since we have a smaller value of $\gamma_c$. 

The mean photon number (MPN) $\langle \hat{a}^{\dagger}\hat{a}\rangle/(\omega_0^2 j)$, is shown in Figure \ref{fig:mean_values_op}(c), where the values are compared with the Dicke model, blue line. 
For $ \gamma_c > 0.5 $ (red line), we obtain lower values than those obtained in the Dicke model. For $ \gamma_c = 0.5 $ (green line), the values are higher due to the contribution of $ \langle \hat {p}_c \rangle $. For $ \gamma_c <0.5 $ (black line), much larger values are obtained since we have a smaller $\gamma_c$. For the chosen values for $ K_1 $ and $ K_2 $, the MPN has been multiplied by a value of $10^{-2}$, to allow a visual comparison with the Dicke model. 
As mentioned above, these high values of MPN  are caused by
the pumping of the non-linear medium, where the effect of the parametric amplification is obtained.  It is a great experimental advantage since it could be possible to achieve the quantum phase transition with far smaller
atom-field couplings than in Dicke model. In this sense, the manipulation of the parameters $(K_1, K_2)$ allows 
to reach different values for both the critical coupling $\gamma_c$ and the MPN,  depending on the experimental needs. 
Finally, the mean value of the population inversion operator is shown in the Figure \ref{fig:mean_values_op}(d). In the normal phase all atoms are, on average, in their ground state and therefore $\langle \hat{J}_z\rangle = -1$. For $\Gamma > 1$, we see that a macroscopic atomic population appears in the system, increasing with respect to $\Gamma$ up to $\langle \hat{J}_z\rangle = 0$, which coincides with the standard Dicke model because the inclusion of nonlinear terms does not affect the atomic subsystem.
\section{Berry Phase Induced by the Cavity}

Our goal in this section 
is to investigate the Berry phase induced by the cavity field and its connection to the QPT present in the extended model, at the thermodynamic limit. The Berry phase is a quantum phase of topological origin acquired, in addition to the dynamical phase, by the eigenstates of a Hamiltonian which are varied
cyclically and adiabatically along a  closed path $\mathcal{C}$ in the parameter space of the system. 
Along a quantum phase transition a non-analyticity appears in the geometric phase of the ground state \cite{hamma2006berry}.
The geometric phase can be found when  a family of Hamiltonians is generated through the application of the unitary transformation on $\hat{H}_{ED}$. This is done by making an adiabatic rotation of the system around the $z$-axis, through of the unitary transformation given by $U(\beta)=e^{-i \beta\hat{a}^{\dagger}\hat{a}}$, and adiabatically varying the angle $\beta$ from $0$ to $2 \pi$, forming a closed path $\mathcal{C}$ in the parameter space. The transformed extended Dicke Hamiltonian is
\begin{align}
    \hat{H}(\beta) = & U(\beta)\hat{H}_D U^{\dagger}(\beta) \nonumber\\
     = &\omega_f \hat{a}^{\dagger}\hat{a}+\omega_0 \hat{J}_z+\frac{\gamma}{\sqrt{N}}(\hat{a}^{\dagger}e^{-i\beta}+\hat{a}e^{i\beta})(\hat{J}_++\hat{J}_-)\nonumber\\
    &  +\frac{K_1}{2}(\hat{a}^{\dagger \ 2}e^{-2i\beta}+\hat{a}^{2}e^{2i\beta})\nonumber\\
    & +i\frac{K_2}{2}(\hat{a}^{\dagger \ 2}e^{-2i\beta}-\hat{a}^{2}e^{2i\beta}).
\end{align}
The effect of the transformation on the Hamiltonian is the addition of a
phase on the creation and annihilation operators, such that
$\hat{a}^{\dagger}\rightarrow \hat{a}^{\dagger}e^{-i\beta} $ and $\hat{a}\rightarrow \hat{a}e^{i\beta} $. The transformed Glauber coherent state for the Eq. \ref{Glauber_CS} is
\begin{align}
    |\alpha(\beta)\rangle = & e^{-|\alpha|^2/2} e^{\alpha \hat{a}^{\dagger}e^{-i\beta}}|0\rangle,
\end{align}
The Berry phase for the ground state is written as
\begin{align}
    \lambda_0 = & i \int_0^{2\pi} \left\langle  \psi_0(\beta)\bigg{|}\frac{d}{d\beta}\bigg{|}\psi_0(\beta)\right \rangle d\beta \nonumber\\
    \nonumber \\
    = & 2\pi\langle \hat{a}^{\dagger}\hat{a}\rangle,
\end{align}
where $|\psi_0(\beta)\rangle=|\alpha(\beta)\rangle \otimes |z\rangle$ is the transformed ground state of the system. This result 
exhibits
the proportionality between the Berry phase and the average number of photons, it has also been found in \cite{chen2006critical}. Therefore, the geometric phase, in the thermodynamic limit, is
\begin{equation}
    \frac{\lambda_0}{\omega_0^2N} = \left\{ \begin{array}{cc}
    0 & \Gamma \leq 1,  \\
    \frac{\pi\Gamma^2(1-\Gamma^{-4})}{8\gamma_c^2}\left[1+\frac{K_2^2}{(\omega-K_1)^2}\right]    &  \Gamma > 1.
    \end{array}
    \right.
    \label{Berry_phase}
\end{equation}
In the normal phase, the geometric phase is zero. In the superradiant phase and for $K_1 = K_2 = 0$ 
Eq. \ref{Berry_phase} reproduce the result for the Dicke model reported in \cite{li2013quantum}.

The first-order derivative of the Berry phase with respect to $\Gamma$, for each phase, is given 
\begin{equation}
    \frac{\partial \lambda_0/(\omega_0^2 N)}{\partial \Gamma} = \left\{ \begin{array}{cc}
    0 & \Gamma \leq 1,  \\
    \frac{\pi \Gamma(1+\Gamma^{-4})}{4\gamma_c^2}\left[1+\frac{K_2^2}{(\omega_f-K_1)^2}\right]    &  \Gamma > 1.
    \end{array}
    \right.
    \label{Berry_phase_derivative}
\end{equation}
\begin{figure}[h]
\centering
\includegraphics[width=1\columnwidth]{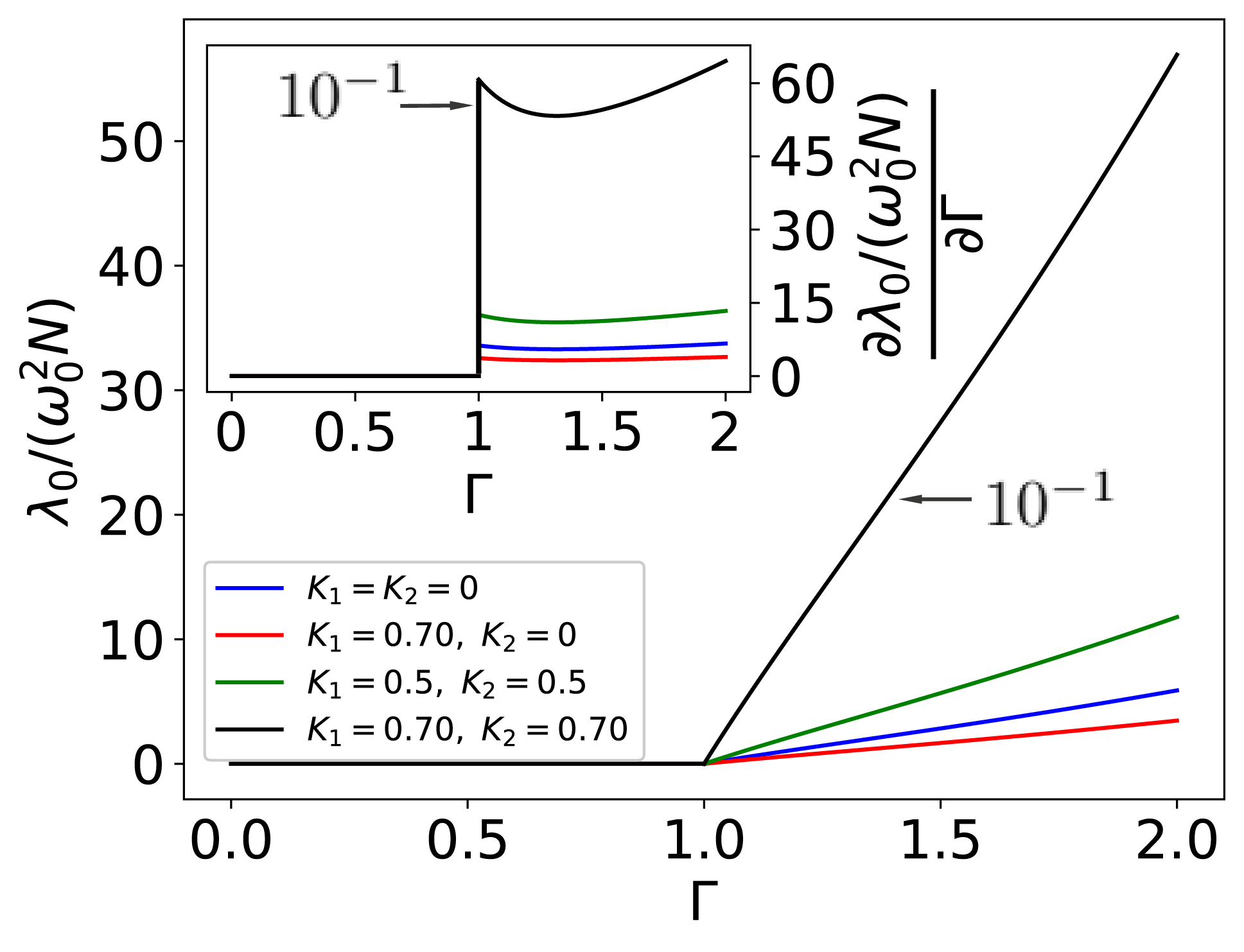}
\caption{The scaled Berry phase $\gamma_0/(\omega_0^2N)$ for the ground state versus the $\Gamma$. In the inset is shown the first-order derivative as a function of $\Gamma$. The values for $(K_1 = K_2=0.7)$ have been divided by
10, and $\omega_f = 1$.}
\label{plot:berry_phase_der}
\end{figure}
In the Fig. (\ref{plot:berry_phase_der}), we show the scaled Berry phase and its first-order derivative as a function of $\Gamma$, for different values of $(K_1,K_2)$. In the normal phase, the Berry phase is zero for any value of $(K_1,K_2)$. For $\Gamma > 1$, we see that the Berry phase increases as $ \Gamma $ increases, in the same way as the average number of photons due to the proportionality given in Eq. \ref{Berry_phase}. The inset shows the first-order derivative with respect to $\Gamma$, with a non-analyticity in its derivative
at the critical point $\Gamma=1$. The values for the corresponding plots for $K_1, K_2 = 0.7$ (black line), are divided by a factor of $10$. In this way, the Berry phase acts as an indicator of quantum phase transitions.

In the thermodynamic limit, the scaling behavior of the Berry phase for the ground state, close to the critical value, $\gamma_c$  $(\Gamma = 1)$, is given by
\begin{align}
    \frac{\gamma_0}{N}(\Gamma \rightarrow 1) = \frac{\pi \omega_0^2}{2\gamma_c}\left[1+\frac{K_2^2}{(\omega_f-K_1)^2}\right]|\Gamma - 1|.
\end{align}
The first order derivative with respect to $ \Gamma$ diverges linearly with $N$, in the vicinity of $\Gamma=1$, in the form
\begin{align}
    \lim_{N \to \infty} \frac{d\gamma_0}{d\Gamma} \bigg|_{\Gamma\rightarrow 1}= \frac{\pi \omega_0^2}{2\gamma_c}\left[1+\frac{K_2^2}{(\omega_f-K_1)^2}\right]N. 
\end{align}
\section{Conclusions}

We have analyzed an extended model of the Dicke model, including two non-linear terms that represent the real and imaginary part of the square of the field amplitude. Through a semiclassical analysis, in the thermodynamic limit, we studied the ground state energy, which exhibits a quantum phase transition, whose dependence in the Hamiltonian parameters was analized in detail.  
Furthermore, our model shows a degenerate parametric amplification effect, which is revealed in the expectation values of the photon field operators and 
has a strong sensitivity to the values of the non-linear parameters, without affecting the atomic sector.
It should be underlined that this effect can be relevant,
since it could allow the experimental access to the ultra-strong coupling regime, reducing the required intensity of the atom-field coupling parameter.

On the other hand, the Berry phase shows a non-analyticity in the ground state for the critical value of the atom-field coupling, confirming
the Berry phase as an indicator of quantum criticality. Furthermore, we observe that the geometric phase scales linearly with $N$ (number of atoms) in the vicinity of the QPT. The geometric phase induced by the cavity field is proportional to the MPN. As a consequence, the high MPN values could help to detected experimentally the Berry phase.



\section{Acknowledgements}
\noindent This works was supported by
and the Departamento Administrativo de Ciencia, Tecnología e Innovación (COLCIENCIAS) of Colombia ``Beca de Doctorados nacionales,
convocatoria 647". CAEG acknowledges the partial support by the Universidad de Antioquia, Colombia, under initiative CODI ES84180154, Estrategia de Sostenibilidad del Grupo de Física Atómica y Molecular, and projects CODI-251594 and 2019-24770. JGH acknowledge partial economical support from DGAPA-PAPIIT project IN104020.

\clearpage

\end{document}